\documentclass[prl,9pt]{revtex4}

\usepackage{amsmath,amssymb}
\usepackage{verbatim}
\usepackage{graphicx}
\usepackage{hyperref}
\usepackage{color}
\usepackage{footnote}
\usepackage{mathtools}

\DeclareFontFamily{OT1}{rsfs}{}
\DeclareFontShape{OT1}{rsfs}{m}{n}{ <-7> rsfs5 <7-10> rsfs7 <10->rsfs10}{} 
\DeclareMathAlphabet{\mycal}{OT1}{rsfs}{m}{n}

\newcounter{rowcount}
\begin{document}

\title{Conformal gravity 1-loop partition function}

\author{Maria Irakleidou}
\email{irakleidou@hep.itp.tuwien.ac.at}
\affiliation{Institute for Theoretical Physics, Technische Universit\"at Wien, Wiedner Hauptstrasse 8--10/136, A-1040 Vienna, Austria}

\author{Iva Lovrekovi\'c}
\email{lovrekovic@hep.itp.tuwien.ac.at}
\affiliation{Institute for Theoretical Physics, Technische Universit\"at Wien, Wiedner Hauptstrasse 8--10/136, A-1040 Vienna, Austria}

\date{\today}


\begin{abstract} 

We evaluate the 1-loop partition function of conformal gravity in four dimensions around an $AdS_4$ background, using the heat kernel techniques. We give expressions for the relevant thermodynamical quantities and compare our results with the ones from the literature.
 
\end{abstract}

\maketitle

\tableofcontents

\section{1. Introduction}

Conformal gravity is an interesting theory of gravity because of its power-counting  renormalizability \cite{Stelle:1976gc,Adler:1982ri}.
Nevertheless it contains ghosts and it should not be considered as a fundamental theory of (quantum) gravity before that issue is resolved. Maldacena showed \cite{Maldacena:2011mk} that one can obtain Einstein gravity from conformal gravity by imposing suitable boundary conditions and t'Hooft pointed out that conformal symmetry may be crucial for understanding the physics at the Planck scale \cite{Hooft:2014daa, Hooft:2009ms,Hooft:2010ac,Hooft:2010nc}. 
It arises from twistor string theory and as a counterterm from five dimensional Einstein gravity \cite{Berkovits:2004jj,Liu:1998bu,Balasubramanian:2000pq}.  On the phenomenological grounds, it was studied by Mannheim considering galactic rotation curves without addition of dark matter \cite{Mannheim:2010xw,Mannheim:2011ds,Mannheim:2012qw}, while it has recently been proven to have a well defined variational principle and finite response functions \cite{Grumiller:2013mxa}. In particular, there are two response functions: the first one is analogous to the Brown-York stress energy tensor and the second one is the partially massless response (PMR), in the sense of Deser, Nepomechie and Waldron \cite{Deser:1983mm,Deser:2001pe}.

One of the most important quantities to be studied in the framework of Anti-de Sitter/Conformal Field Theory (AdS/CFT) correspondence is the partition function.  In this framework, the partition function of the gravity theory in $d$ dimensions corresponds to the partition function of the conformal field theory in $d-1$ dimensions. 
For example, such studies were considered from the perspective of the canonical partition function evaluation for non-interacting conformal higher spin in \cite{Beccaria:2014jxa} using the operator counting method.
In the context of gauge/gravity correspondence it was mostly studied in lower dimensions  \cite{Maloney:2007ud,Gaberdiel:2010xv,Bertin:2011jk,David:2009xg}.
One method to obtain the partition function is the heat kernel method \cite{Vassilevich:2003xt} which can be used for particular backgrounds and operators.
On certain symmetric backgrounds such as sphere or hyperbolic space the heat kernel can be analytically evaluated. 
When an analytic computation is not convenient one may consider
studying the heat kernel coefficients. One of the methods for that is the world-line formalism as in \cite{Fliegner:1994zc,Fliegner:1997rk}. In particular, 1-loop Einstein gravity with matter has been studied using the world line formalism on general backgrounds \cite{Bastianelli:2013tsa} where its representation via world line path integrals  \cite{Fliegner:1997rk,Fliegner:1994zc} has proven to give correct 1-loop divergencies \cite{Bastianelli:2013tsa}.
Some of the operators for which the partition function has been evaluated are the Laplace operator \cite{Camporesi:1994ga,Gopakumar:2011qs,Beccaria:2016tqy}, the conformal higher spin operator \cite{Beccaria:2014jxa,Beccaria:2016tqy} and, GJMS operators \cite{Beccaria:2016tqy,Beccaria:2015vaa}. The method of heat kernel can be also used for the above operators which are more general, as well as for the differential operator whose construction is important in conformal differential geometry in four dimensions, i.e. the Paneitz operator, \cite{Fradkin:1981iu,Fradkin:1981jc,Paneitz:2008aa}.

In this paper we compute and analyze the 1-loop partition function of conformal gravity in four dimensions using the heat kernel method \cite{Vassilevich:2003xt}.  
For this we have to linearize the equations of motion and evaluate the path integral corresponding to the 1-loop partition function. We find it convenient to express the result in terms of the determinants of Laplacian operators around an $AdS_4$ background. 
Then we analytically evaluate the determinants via heat kernel techniques. For that we compute the traced heat kernel by analytic continuation of $S^4$ to $\mathbb{H}^4$. 

We compare our result with the 1-loop partition function for Einstein gravity, pointing out the clear contribution from the ghost mode and the partially massless mode which appear in conformal gravity, however do not appear in Einstein gravity.






This paper is organised in the following way: In chapter 2 we introduce the preliminaries for the computation of the 1-loop partition function, in chapter 3 we fix the gauge and explain the ghost determinant and finally in chapter 4 we introduce the formalism we use for the evaluation of the partition function and the free energy. We summarise our results in chapter 5.



\section{2. Preliminaries}
The conformal gravity action in four dimensions consists of the Weyl tensor $C^{\alpha}{}_{\beta \gamma \delta}$, the completely traceless part of the Riemann tensor, and it is constructed such that it is conformally invariant under a (local) conformal transformation of the metric tensor $g_{\mu\nu}\rightarrow e^{2\sigma} g_{\mu\nu}$.  It takes the form
\begin{equation}
\label{eq:m}
\Gamma=\,\alpha\int_{M}\!\!\! d^{4}x\, \sqrt{|g|}\,\,C_{\alpha}{}^{\beta\gamma\delta} \,C^{\alpha}{}_{\beta\gamma\delta}
\end{equation}
where $\alpha$ is dimensionless coupling constant. The equations of motion are
\begin{equation}
\label{eq:eomm}
B_{\mu\nu}\equiv 2 \nabla^{\beta}\nabla^{\alpha}C_{\alpha \mu\nu\beta} + C_{\alpha\mu\nu\beta} R^{\alpha\beta}=0
\end{equation}
where $B_{\mu\nu}$ is the Bach tensor.\\
We denote the fluctuation $h_{\mu\nu}$ of the metric tensor $g_{\mu\nu}$ around the $AdS_{4}$ background ($\bar{g}_{\mu\nu}$) as
\begin{equation}
\label{eq:perth}
g_{\mu\nu}=\bar{g}_{\mu\nu} + h_{\mu\nu} .
\end{equation}
From now on tensor fields with a bar (\,$\bar{}$\,) denote their value on the $AdS_{4}$ background.
 The second variation of \eqref{eq:m} gives 
\begin{equation}
\label{eq:sv}
\delta^{(2)}\Gamma=\,\alpha\int_{M}\!\!\! d^{4}x\, \sqrt{|\bar{g}|} \Big( \delta B^{\mu\nu}h_{\mu\nu}+\bar{B}^{\mu\nu}\delta h_{\mu\nu}\Big)=\alpha\int_{M}\!\!\! d^{4}x\, \sqrt{\bar{g}}  \delta B^{\mu\nu}h_{\mu\nu}
\end{equation}
when evaluated on-shell for the background metric.
\newline
Additionally, we York-decompose the fluctuation $h_{\mu\nu}$ into a transverse-traceless part ($h_{\mu\nu}^{TT}$), a ``trace" part ($\hat{h}$) and a gauge part ($\bar{\nabla}_{(\mu} \xi_{\nu)}$)
\begin{equation}
\label{eq:york}
h_{\mu\nu}=h^{TT}_{\mu\nu} + \frac{1}{4} \bar{g}_{\mu\nu} \hat{h}+2 \bar{\nabla}_{(\mu} \xi_{\nu)}
\end{equation}
with $\bar{\nabla}^{\mu}h^{TT}_{\mu\nu}=0=\bar{g}^{\mu\nu}h^{TT}_{\mu\nu}$. 
\newline
The second variation \eqref{eq:sv} now takes the form
\begin{align}
\label{eq:sec}
\delta^{2}S&=\int_{M}\!\!\! d^{4}x\, \sqrt{|\bar{g}|}\,\, h^{\mu\nu}_{TT}\Big(\!-8-6 \bar{\nabla}_{\alpha}\bar{\nabla}^{\alpha}-\bar{\nabla}_{\beta}\bar{\nabla}^{\beta}\bar{\nabla}_{\alpha}\bar{\nabla}^{\alpha}\Big) h^{TT}_{\mu\nu} \nonumber\\
&=\int_{M}\!\!\! d^{4}x\, \sqrt{|\bar{g}|}\,\, h^{\mu\nu}_{TT}\,\,L\,\,h^{TT}_{\mu\nu} 
\end{align}
with
\begin{equation}
\label{eq:op}
L\equiv-( 8 + 6 \bar{\nabla}^{2}+\bar{\nabla}^{4}).
\end{equation}
From now on we omit the bar (\,$\bar{}$\,) from the Laplacian operator. This result is consistent with the one obtained in \cite{Giombi:2014yra} and the linearised equations of motion from  \cite{Lu:2011ks} and \cite{Lu:2011zk}. Since the full theory is diffeomorphic and conformally invariant we expect this to hold at the linearized level and this is indeed the case: the operator \eqref{eq:op} does not depend on the vector $\xi_{\mu}$ or the ''trace'' part $\hat{h}$. 
\newline
The 1-loop partition function is determined from the path integral over all (smooth) fluctuations $h_{\mu\nu}$ around a background of the second variation of the action:
\begin{equation}
Z_{1-loop}=\int Dh_{\mu\nu}\times ghost\times \exp(-\delta^{(2)}\Gamma)
\end{equation}
in which ``ghost" refers to the determinants produced via elimination of the gauge degrees of freedom. We will explicitly determine these contributions in the next chapter.

\section{3. Gauge fixing and ghost determinant}
As we already mentioned before, the second variation of the conformal gravity action \eqref{eq:sec} contains neither the vector $\xi_{\mu}$ nor the ``trace'' modes $\hat{h}$ due to conformal invariance. 
Therefore, the functional integration over the gauge and ``trace'' degrees of freedom can be perfomed trivially. They give the volume of the conformal group which is an infinite constant and has to be eliminated. The procedure is done as follows: the path integral measure $Dh_{\mu\nu}$ is divided by the volume of the conformal group which is useful to express in terms of a ghost (Faddeev-Popov) determinant. Then, this ghost determinant is given by the Jacobian of the transformation $h_{\mu\nu} \rightarrow (h_{\mu\nu}^{TT}, \xi_{\mu}, \hat{h})$:
\begin{equation}
\label{eq:path}
Dh_{\mu\nu}=Z_{gh} \,Dh^{TT}_{\mu\nu} \,D\xi_{\mu}\, D\hat{h}
\end{equation}
We require
\begin{equation}
\label{eq:one}
1=\int Dh_{\mu\nu}\,\, \exp\Big[-\!\!\int d^{4}x\sqrt{\bar{g}}\,\, h_{\mu\nu} h^{\mu\nu}\Big]=\int Z_{gh} Dh^{TT}_{\mu\nu} D\xi_{\mu} D\hat{h}\,\, \exp\Big[-\!\!\int d^{4}x\sqrt{\bar{g}}\,\, h_{\mu\nu} (h^{TT}, \xi, \hat{h}) h^{\mu\nu} (h^{TT}, \xi, \hat{h})\Big]
\end{equation} 
Similarly, we recquire for each mode of the York-decomposition:
\begin{align}
1&=\int Dh^{TT}_{\mu\nu}\,\, \exp\Big[-\!\!\int d^{4}x\sqrt{\bar{g}}\,\, h^{TT}_{\mu\nu} h_{TT}^{\mu\nu}\Big] \label{y}\\
1&=\int  D\xi_{\mu}\,\, \exp\Big[-\!\!\int d^{4}x\sqrt{\bar{g}}\,\, \xi^{\mu} \xi_{\mu}\Big] \label{y2}\\
1&=\int  D\hat{h}\,\, \exp\Big[-\!\!\int d^{4}x \sqrt{\bar{g}}\,\,\hat{h}^{2}\Big] \label{y3}
\end{align}
To cancell the mixing between the modes of different types in \eqref{eq:one} it is convenient to decompose the gauge part $\xi_{\mu}$ into a transverse and a scalar part as $\xi_{\mu}=\xi^{T}_{\mu}+\nabla_{\mu} \sigma$, with $\nabla^{\mu}\xi^{T}_{\mu}=0$ and the ``trace" part $\hat{h}$ as $\hat{h}=\tilde{h} - 2 \nabla^{2}\sigma$. The latter transformation gives a unit Jacobian whereas the first one using \eqref{y2} becomes:
\begin{align}
\label{eq:tv}
1&=\int  D\xi^{T}_{\mu} D\sigma J_{1}\,\, \exp\Big[-\!\!\int d^{4}x\sqrt{\bar{g}}\,\, (\xi_{T}^{\mu} \xi^{T}_{\mu}-\sigma \nabla^{2} \sigma)\Big] \nonumber\\
\Rightarrow  J_{1}&=\Big[ \det(-\nabla^{2})_{(0)}\Big]^{\frac{1}{2}}
\end{align}
where the subscript $(0)$ denotes the scalar mode $\sigma$. Then, the decomposition of the inner product \eqref{eq:one} is othogornal 
\begin{equation}
\label{eq:m2}
h_{\mu\nu} h^{\mu\nu} = h^{TT}_{\mu\nu} h_{TT}^{\mu\nu} -2 \xi^{\mu}_{T}(\nabla^{2}-3) \xi^{T}_{\mu}+3 \sigma (-\nabla^{2}) (-\nabla^{2}+4) \sigma+\frac{1}{4}\hat{h}^{2}
\end{equation}
and we find
\begin{align}
\label{eq:finalone}
1&=\int Dh^{TT}_{\mu\nu} D \xi^{T}_{\mu} D \sigma \, J_{2} \exp \Big[ -\int d^{4}x\sqrt{\bar{g}}\,\,\Big( h^{TT}_{\mu\nu} h_{TT}^{\mu\nu} -2 \xi^{\mu}_{T}(\nabla^{2}-3) \xi^{T}_{\mu}+3 \sigma (-\nabla^{2}) (-\nabla^{2}+4)\Big) \Big] \nonumber\\
\Rightarrow J_{2} &= \Big[ \det(-\nabla^{2}+3)^{T}_{(1)}\, \det(-\nabla^{2})_{(0)}\,\det(-\nabla^{2}+4)_{(0)}\Big]^{\frac{1}{2}}
\end{align}
where the notation $^{T}\,_{(1)}$ denotes the transverse vector mode $\xi_{\mu}^{T}$. 
Therefore for \eqref{eq:path} we have
\begin{align}
\label{eq:path2}
Dh_{\mu\nu}&= J_{2} \, Dh^{TT}_{\mu\nu}\, D\xi_{\mu} \,D \sigma D\hat{h}= \frac{J_{2}}{J_{1}} \, Dh^{TT}_{\mu\nu}\,D \xi_{\mu} \,D \hat{h} \nonumber\\
\Rightarrow Z_{gh} &=\frac{J_{2}}{J_{1}}=\Big[ \det(-\nabla^{2}+3)^{T}_{(1)}\det(-\nabla^{2}+4)_{(0)}\Big]^{\frac{1}{2}}
\end{align}
We now have all the necessary ingredients to calculate the 1-loop partition function and we proceed in the next chapter.

\section{4. 1-loop partition function}
The 1-loop partition function takes the form
\begin{align}
\label{eq:loop}
Z_{1-\text{loop}}&=\int Dh_{\mu\nu}\,[V_{\text{diff}}]^{-1}\,[V_{\text{conf}}]^{-1}\,\, exp [-\delta^{2} \Gamma  ] \nonumber\\
&=Z_{\text{gh}}\int  Dh^{TT}_{\mu\nu}\,\, \exp\Big[-\!\!\int_{M}\!\!\! d^{4}x\, \sqrt{\bar{g}}\,\, h^{\mu\nu}_{TT}\,( - 8 - 6 \nabla^{2}-\nabla^{4})\,h^{TT}_{\mu\nu} \Big] \nonumber\\
&= \Big[ \frac{\det(-\nabla^{2}+3)^{T}_{(1)}\det(-\nabla^{2}+4)_{(0)}}{\det(-\nabla^{2}-4)_{(2)}^{TT} \det(-\nabla^{2}-2)_{(2)}^{TT}}  \Big]^{\frac{1}{2}} \nonumber\\
&=Z_{(1)}\,Z_{(0)}\,Z_{(2),4}^{-1}\,Z_{(2),2}^{-1}
\end{align}
where $Z_{(S)}$ is the partition function of the relevant modes of spin $S=0,1,2$ and $Z_{(2),r}$ is the partition function of the spin-$2$ modes with constant terms $r=4, 2$. This result agrees exactly with the partition function obtained in \cite{Beccaria:2014jxa}, equation (3.16). This was as well considered in references therein, namely \cite{Fradkin:1985am} and \cite{Fradkin:1983zz}. The spin-0 and the spin-1 modes that appear in the numerator of the above expression are pure gauge and the dynamical degrees of freedom are the two spin-2 transvere-traceless modes that appear in the denumerator: the determinant $(-\nabla^{2}-2)_{(2)}^{TT}$ corresponds to the spin-2 massless graviton and the determinant $(-\nabla^{2}-4)_{(2)}^{TT}$ corresponds to the the spin-2 partially massless graviton (PMG). 
At this point, it is useful to compare our result with the 1-loop partition of Einstein gravity (around the $AdS_{4}$ background):
{\setlength\arraycolsep{0.01pt}
\begin{eqnarray}
\label{eq:eg}
Z_{\text{Einstein}, 1-\text{loop}}&=&\Big[\frac{\det(-\nabla^{2}+3)^{T}_{(1)}}{\det(-\nabla^{2}-2)_{(2)}^{TT}}\Big]^{\frac{1}{2}} \label{egrav1}\\
\Rightarrow Z_{\text{CG}, 1-\text{loop}}&=&Z_{\text{Einstein}, 1-\text{loop}}\,\Big[\frac{\det(-\nabla^{2}+4)_{(0)}}{\det(-\nabla^{2}-4)^{TT}_{(2)}}\Big]^{\frac{1}{2}}
\end{eqnarray}}
We clearly see now the additional degrees of freedom that appear in CG in comparison to the Einstein case, namely the scalar part (that is pure gauge and corresponds to the Weyl invariance) and the PMG.
\newline
\newline
We continue now to evaluate analytically these determinants using the heat kernel methods and the prescription of \cite{Gaberdiel:2010xv}. 
The partition function and the determinant of an operator are related to the trace of the heat kernel $K^{(S)}(t)$ via
\begin{equation}
\label{eq:pf}
\ln Z_{(S)}=-\frac{1}{2}\ln \det(-\nabla^{2}+m^{2})_{(S)}=-\frac{1}{2} \text{Tr} \ln(-\nabla^{2}+m^{2})_{(S)}=-\frac{1}{2}\int\limits_0^{\infty}\frac{dt}{t}K^{(S)}(t)
\end{equation}
where the traced heat kernel is defined as
\begin{equation}
\label{eq:trhkgen}
K^{(S)}(t)\equiv Tr \exp \Big[\!-t(-\nabla^{2}+m^{2})_{(S)}\Big]
\end{equation}
and $m$ is a constant. The determinant of symmetric transverse traceless operators (STT) of arbitrary spin $S$ can be computed using the method described in \cite{Gopakumar:2011qs} which considers the heat kernel on $AdS_{2n+1}$ backgrounds, i.e. odd-dimensional hyperboloids and in \cite{Lovrekovic:2015thw}, which considers the heat kernel of STT tensors on $AdS_{2n}$, i.e. even-dimensional hyperboloids. In general however, one has to keep in mind that the case of even-dimensional hyperboloids is more subtle. We shall briefly recollect the main results of the heat kernel techniques on even-dimensional spaces specialising in the case of $AdS_4$.

\subsection{The traced heat kernel on thermal quotient of $AdS_{4}$}

To find the traced heat kernel on thermal quotient of $AdS_{4}$ we consider the quotient space $\mathbb{H}^4\simeq SO(4,1)/SO(4)$ obtained by analytic continuation from the $4$-sphere $S^{4}\simeq SO(5)/SO(4)$. The traced heat kernel coefficient is then given \cite{Gopakumar:2011qs} by
\begin{equation}
\label{eq:trace} 
K^{(S)}(t)=\frac{\beta}{2\pi}\sum\limits_{k\in\mathbb{Z}}\sum\limits_{\vec{m}}\int\limits_{0}^{\infty}d\lambda \,\chi_{\lambda,\vec{m}}(\gamma^k)\,e^{{t E_R^{(S)}}}
\end{equation}
where  $E_R^{(S)}$ are the eigenvalues of the spin S Laplacian operator on the quotient space $\mathbb{H}^4$, $\chi_{\lambda,\vec{m}}(\gamma^k)$ is the Harish-Chandra character in the principal series of $SO(4,1)$,  $\gamma$ is an element of the thermal quotient of $S^4$, $\beta $ is the inverse temperature and $(\lambda,\vec{m})$ denotes the principal series representation \cite{Gopakumar:2011qs}. We will now outline the procedure to find the eigenvalues in the case of the symmetric transverse traceless tensors (STT) that we are interested in, whereas the explicit derivation of $\gamma$ is given in the appendix.
\newline
The unitary irreducible representations of $SO(4,1)$, denoted as $R$, are characterised via the array 
\begin{align}
\begin{array}{ccc}
R=(m_{1}, m_2)=(i\lambda, m_2) & \,\,\, \text{       with     }\,\,\, & \lambda \in \mathbb{R} \label{ac1}
\end{array}
\end{align}
where $m_{2}$ is a non-negative (half-)integer while the unitary irreducible representations of $SO(4)$, denoted as $S$, are characterised by the array
\begin{align}
\begin{array}{ccc}
S=(s_1, s_2) & \,\,\, \text{ with } \,\,\, &s_1\geq s_2\geq0 \label{ac2}
\end{array}
\end{align}
where $s_1, s_2$ are (half-) integers. The eigenvalues of the spin S Laplacian operator in the quotient space $SO(4,1)/SO(4)$ are given by
\begin{equation}
\label{eq:eig}
-E_{R}^{(S)}=C_{2}(R)-C_{2}(S)
\end{equation}
where $C_{2}(R)$ and $C_{2}(S)$ are the quadratic Casimirs for the unitary representations of $SO(4,1)$ and $SO(4)$ respectively:
\begin{equation}
\label{eq:eg}
C_{2}(R)=m \cdot m +2 r _{SO(4,1)} \cdot m  \,,\,\,\,\,\,\,\,\,\,\,\,\,\,\,\,C_{2}(S)=s \cdot s+2r _{SO(4)} \cdot s
\end{equation}
Here the dot product is the usual Euclidean one and $r_{i, SO(4,1)}=\frac{5}{2}-i$, $r_{i, SO(4)}=2-i$ and $i=1,2,3$. We find that
\begin{equation}
\label{eq:eigen2}
-E_{R}^{(S)}=\lambda^{2}+\frac{9}{4}+S
\end{equation}
The Harish-Chandra character in the principal series of SO(4,1) \cite{hirai} is
\begin{equation}
\chi_{\lambda,\vec{m}}(\beta, \phi_{1})=\frac{(e^{-i\beta\lambda}+e^{i\beta\lambda})\chi_{\vec{m}}^{SO(3)}(\phi_1)}{e^{-\frac{3\beta}{2}}|e^{\beta}-1| |e^{\beta}-e^{i\phi_1}|^2} 
\end{equation}
where $\chi_{\vec{m}}^{SO(3)}(\phi_1)$ is the character of the representation of $SO(3)$. For the thermal quotient we are considering $\beta\neq0$, $\phi_1=0$ and $\chi_{\vec{m}}^{SO(3)}(0)=1+2S$ \cite{Dolan:2005wy} and therefore the above expression simplifies to
\begin{equation}
\chi(\beta,\phi_1)=(1+2 S)\frac{\cos(\beta\lambda)}{4\sinh^{3}\left(\frac{\beta}{2}\right)}
\end{equation}

The traced heat kernel \eqref{eq:trace} finally takes the form
\begin{equation}
K^{(S)}(t)=\frac{\beta(1+2 S)}{8\sqrt{\pi t}}\sum\limits_{k\in\mathbb{Z}_+}\frac{1}{\sinh^{3}\frac{k\beta}{2}}e^{-\frac{k^2\beta^2}{4t}-t(\lambda^2+S)}
\end{equation}
Notice that the summation goes over $\mathbb{Z}_+$ because we do not include $k=0$ since it diverges. It appears since the volume of $AdS$ is infinite, over which the coincident heat kernel on the full $AdS_{4}$ is integrated. That term does not depend on $\beta$ and it can be absorbed  into the parameters of the gravity theory.
Now, we have all the ingredients to evaluate the partition function for arbitrary spin S \eqref{eq:pf} 
\begin{equation}
\label{eq:pf3}
\ln Z_{(S)}=-(1+2S)\sum\limits_{k\in\mathbb{Z}_+}\frac{e^{-k\beta\big(\frac{3}{2}+\sqrt{\frac{9}{4}+m^{2}+S}\big)}}{(1-e^{-k\beta})^{3}k}.
\end{equation}
From the above expression and \eqref{egrav1} we can read out the Einstein part contribution, the PMG and the scalar field. The 1-loop partition function for a scalar field on $AdS_{4}$ background matches the expression obtained in \cite{Gopakumar:2011qs,Gibbons:2006ij} and the one obtained from Hamiltonian analysis \cite{Gibbons:2006ij}. This is
\begin{equation}
\ln Z_{(0)}=\sum_k^{\infty}\frac{-1}{k(1-e^{-k\beta})^3}e^{-k\beta\left(\frac{-3}{2}\right)}\left( e^{-k\beta\sqrt{\frac{9}{4}+4}} \right) \label{zcg}
\end{equation}

Therefore now we can evaluate each spin contribution in the 1-loop partition function \eqref{eq:loop} and our final and main result is
{\setlength\arraycolsep{0.01pt}
\begin{eqnarray}
\label{eq:part}
\ln Z_{1-\text{loop}}&=&-\sum\limits_{k\in\mathbb{Z}_+}\frac{e^{-2 k \beta} (-5 + 4 e^{-2 k \beta} -5 e^{-k \beta})}{(1- e^{-k \beta})^{3} k} \nonumber\\
&=&-\sum\limits_{k\in\mathbb{Z}_+}\frac{q^{2k} (-5 + 4 q^{2k} -5 q^{k})}{(1- q^{k})^{3} k} 
\end{eqnarray}}where $q=e^{-\beta}$.
Interestingly, it is proportional to the 1-loop partition function evaluated on $S^1\times S^3$ background and $\mathbb{R}^4$ background in \cite{Beccaria:2014jxa}. The possible explanation is in the relation of the two backgrounds with the conformal factor and the fact that the conformal spin-2 field is Weyl invariant. 
\newline

\subsection{Thermodynamic quantities}
From (\ref{eq:part}) one can immediately infer the 1-loop (Helmholtz) free energy $(-\frac{1}{\beta}\ln Z_{1-\text{loop}})$:
\begin{equation}
\label{eq:fe}
F_{1-\text{loop}}=\sum\limits_{k\in\mathbb{Z}_+}\frac{e^{-2 k \beta} (-5 + 4 e^{-2 k \beta} -5 e^{-k \beta})}{(1- e^{-k \beta})^{3} k \beta}
\end{equation}
in the literature often considered multiplied with $\beta$.  
Note that the classical contribution to free energy is known to vanish on the $AdS_4$ background since the Weyl tensor vanishes and the Euclidean action (\ref{eq:m}) does not have to be renormalised \cite{Grumiller:2013mxa}.

 The Einstein part of the partition function we can compare to the free energy in conventions of \cite{Giombi:2014yra} taking the spin two case. Its 1-loop partition function that we obtain
\begin{eqnarray}
\label{eq:eq}
-\beta F_{\text{EG}_{1-\text{loop}}}= lnZ_{\text{EG}_{1-\text{loop}}}&=&-\sum\limits_{k\in\mathbb{Z}_+}\frac{q^{3k} (-5 +3 q^{k})}{(1- q^{k})^{3} k} 
\end{eqnarray}
 agrees with the one from \cite{Giombi:2014yra}. 
 
 
The 1-loop contribution to the entropy $(S_{1-\text{loop}}=-\frac{\partial{F_{1-\text{loop}}}}{\partial{T}})$ 
\begin{equation}
\label{eq:entropy}
S_{1-\text{loop}}=\sum\limits_{k\in\mathbb{Z}_+}\frac{e^{\frac{-k}{T}}\Big(20\, k\, e^{\frac{2 k}{T}}+ 4 (k+T)+ 5\,(2k+T)\, e^{\frac{3 k}{T}}  - (16k+9T)\,e^{\frac{k}{T}}\Big)}{k T (-1+e^{\frac{k}{T}})^{4}}
\end{equation}
is non-divergent, where $T$ is the temperature.
One might be tempted to give a physical interpretation to the non-divergent 1-loop contribution to the entropy of the $AdS_{4}$ but nevertheless one should be careful and question the validity of the semi-classical approximation for this solution: the classical contribution is not negligible neither is the 1-loop part. 
 Therefore, one in general should expect full quantum corrections to contribute to these terms.

\section{5. Summary}

In the article we compute the 1-loop partition function of conformal gravity and analyse it: it consists of the scalar, vector and tensor determinant with non-vanishing mass. The determinants of the higher spin operators computed on the thermal AdS and the partition function of a scalar determinant obtained via procedure of analytic continuation agree with ours, and the partition function from \cite{Beccaria:2014jxa} obtained using operator counting method.
Operator counting method used in \cite{Beccaria:2014jxa} has proven to give the partition function for the higher spin symmetric transverse traceless operators which are again in agreement with the ones obtained using the group theoretic approach of evaluation of heat kernel on the thermal AdS that we use here.
Our method of evaluation of heat kernel on thermal $AdS_{4}$ shows agreement when we consider determinants coming purely from the linearised eom. Our main result  is consideration of the pure conformal gravity action in four dimensions in which we can clearly differ and separate the partition function into the one coming from Einstein gravity, part specific for the conformal gravity, contribution from the partially massless mode and the part originating from the conformal ghost, resulting in the scalar determinant. 

We can compare the structure of the partition function with the structure of the partition function of conformally invariant theory in three dimensions, which is Chern-Simons gravity. One can notice that the partition function constitutes of the analogous terms as conformal gravity in four dimensions. Partition function of Einstein Gravity (EG), contribution form conformal ghost and partially massless contribution \cite{Bertin:2011jk}. The analogous structure is found if we compare the partition function to the partition function of conformal gravity in six dimensions. Partition function in that case consists of the part that corresponds to Einstein gravity in six dimensions, contribution from the conformal ghost and partially massless mode as well. However, there is one more determinant that contains mass which appears due to the fact that in that case one obtains six derivative theory \cite{Lovrekovic:2015thw}.

It is important to stress that our result is comparable to the partition function computed on the $S^1\times S^{3}$ and $\mathbb{R}^4$ background that has been compared to the partition function computed on the field theory side.


The further possible application of 1-loop partition function would be to consider the full partition function of conformal gravity and computation of the 1-loop conformal gravity keeping an arbitrary background metric that would allow to write the 1-loop CG using world line path integrals \cite{Fliegner:1994zc,Fliegner:1997rk}. Since in the present case not all classical solutions to conformal gravity are known, before addressing the full partition function one has to list all the classical solutions of the theory and study it perturbatively  around a classical solution. While in case of three dimensional gravity, one can consider the full partition function while classical solutions are known \cite{Maloney:2007ud}.
\newline
\newline
\textbf{Acknowledgments}
\newline
We are grateful to Daniel Grumiller and Dmitri Vassilevich for the guidance and help throughout this project and thank Arkady Tseytlin for very useful discussions.  We also thank Stefan Prohazka, Jakob Salzer, Friedrich Sch\"oller, Florian Preis and Stefan Stricker for the discussions and Matteo Beccaria for communications. M. I. was supported by the FWF Project "Doctoral program Particles and Interactions"  DKW1252-N27 and the FWF project I 1030-N16. I. L. was funded by the START project Y 435-N16 of the Austrian Science Fund (FWF), the FWF project I 952-N16 and {\it Forschungsstipendien 2015}  of {\it Technische Universit\"at Wien}.

\section{A. Appendix}


To define a thermal section on the sphere $S^{4}$ we use appropriately adjusted geometrical construction from \cite{Gopakumar:2011qs}.  
The definition of the  element $\gamma$ we give here is described in terms of the thermal section on the sphere $S^{4}\simeq SO(5)/SO(4)$. Using the analytic continuation described in \cite{Gopakumar:2011qs} and \cite{Lovrekovic:2015thw} one obtains the analogous quantity on the hyperbolic space $\mathbb{H}^{4}\simeq SO(4,1)/SO(4)$ that corresponds to Euclidean $AdS_4$.

With two compact Lie groups SO(5) and SO(4) one can define the coset space SO(5)/SO(4) obtained via right action of elements g of SO(5) such that $SO(5)/SO(4)={g SO(4)}$. Where SO(5) is a principal bundle over SO(5)/SO(4) with fibre isomorphic to SO(4) and $\pi$ a projection map $\pi(g)=g SO(4)$  $\forall$ $g\in SO(4)$ and section $\sigma(x)$ such that $\sigma:SO(5)/SO(4)\rightarrow SO(5)$. Where $\pi \circ e=0$ and $e$ is identity in $g SO(4)$.

One can define then the heat kernel on the quotient space $\Gamma\backslash SO(5)/SO(4)$ where $\Gamma\simeq\mathbb{Z}_{4}$ for the thermal quotient on the $4-sphere$ that can be embedded into SO(5).  For evaluation of the thermal section compatible with quotienting $\Gamma$ one defines $\gamma\in\Gamma$ that acts on points $x=g SO(4)\in SO(5)/SO(4)$ where $\gamma: g SO(4)\rightarrow \gamma \cdot g SO(4)$. $\sigma(x)$  is then compatible with quotienting $\Gamma$ if \begin{equation}\sigma(\gamma(x))=\gamma \cdot\sigma(x).\label{ts}\end{equation}

Explicitly, we express the thermal quotient using "triple-polar" coordinates on $S^4$, that are complex numbers $z_1,z_2,z_3$ which satisfy the condition 
\begin{equation}
|z_1|^2+|z_2|^2+|z_3|^2=1
\end{equation}
The quotient is
\begin{equation}
\gamma: \{\phi_i\}\rightarrow\{\phi_i+\alpha_i\}\label{tq}
\end{equation}
with $\phi_1,\phi_2$ phases of the z's and $n_i\alpha_1=2\pi$ for some $n_i\in\mathbb{Z}$ where not all $n_i$s are simultaneously zero.  (For the thermal quotient here, we need $\alpha_i=0 (\forall i\neq1) $). To embed $\Gamma$ in SO(5) we decompose complex numbers into 5 coordinates that are real and embed the $S^4$ in $\mathbb{R}^5$. 
\begin{align}
x_1&=\cos\theta\cos\phi_1 & x_2&=\cos\theta\sin\phi_1 \\
x_3&=\sin\theta\cos\psi\cos\phi_2 & x_4&=\sin\theta\cos\psi\sin\phi_2 \\
x_5&=\sin\theta\sin\psi & 
\end{align}
The coset representative in $SO(5)$ is constructed for the point x with the above coordinates. Denote the north pole, point in $R^5$, with (1,0,0,0,0) and construct a matrix g(x) that rotates it to  the generic point x. g(x) is constructed so that $g(x)\in SO(5)$ contains point x on $S^4$ so that there is one to one correspondence between them unto multiplication by an element of SO(4) that keeps north pole invariant. That matrix g(x) can be
\begin{equation}
g(x)=e^{i\phi_1Q_{12}}e^{i\phi_2Q_{34}}e^{i\psi Q_{35}}e^{i\theta Q_{13}}
\end{equation}
and the Qs are generators of SO(5). That can be recognised as an instance of a section in G over G/H. We can write the  action of the thermal quotient (\ref{tq})
on $g(x)$ in a form of embedding of $\Gamma$ in SO(5)
\begin{equation}
\gamma : g(x)\rightarrow g(\gamma(x))=e^{i\alpha_1Q_{12}}e^{i\alpha_2Q_{34}}\cdot x=\gamma\cdot g(x)
\end{equation}
where $\cdot$ is matrix multiplication. Since we can as well recognise the property (\ref{ts})  we can choose the thermal section as  
\begin{equation}
\sigma_{th}(x)=g(x).
\end{equation}

To construct the heat kernel on $\Gamma\backslash SO(5)/SO(4)$ one uses method of images in a sense of \cite{Gopakumar:2011qs}, \cite{David:2009xg} and continuation to a hyperbolic space $\mathbb{H}_4$. This way one obtains the expression for the thermal AdS with $\Gamma\simeq \mathbb{Z}$, not $\mathbb{Z}_4$, as for the sphere. This result for the heat kernel corresponds to (\ref{eq:trace}) from the main text.


\bibliography{bibliothek4Dcg}

\end{document}